%% file: template.tex
\documentclass{vgtc}                          % final (conference style)
\graphicspath{{figures/}{pictures/}{images/}{./}} % where to search for the images

\usepackage{times}                     % we use Times as the main font
\usepackage{tabu}                      % only used for the table example
\usepackage{booktabs}                  % only used for the table example
\usepackage{lipsum}                    % used to generate placeholder text
\usepackage{mwe}                       % used to generate placeholder figures

\usepackage{mathptmx}                  % use matching math font
\input{command}

\onlineid{1124}

\vgtccategory{Research}

\vgtcinsertpkg

\title{Enhancing XAI Interpretation through a Reverse Mapping from Insights to Visualizations}

\author{Aniket Nuthalapati\thanks{The first two authors made equal contributions.}\ \ \thanks{e-mail: nutha010@umn.edu} \\%
        \scriptsize University of Minnesota %
\and Nicholas Hinds\footnotemark[1]\ \ \thanks{e-mail: hinds084@umn.edu}\\ %
     \scriptsize University of Minnesota %
\and Brian Y. Lim\thanks{e-mail: brianlim@comp.nus.edu.sg}\\ %
     \scriptsize National University of Singapore %
\and Qianwen Wang\thanks{e-mail: qianwen@umn.edu}\\ %
    \scriptsize University of Minnesota %
     }

\teaser{
  \centering
  \includegraphics[width=0.85\linewidth]{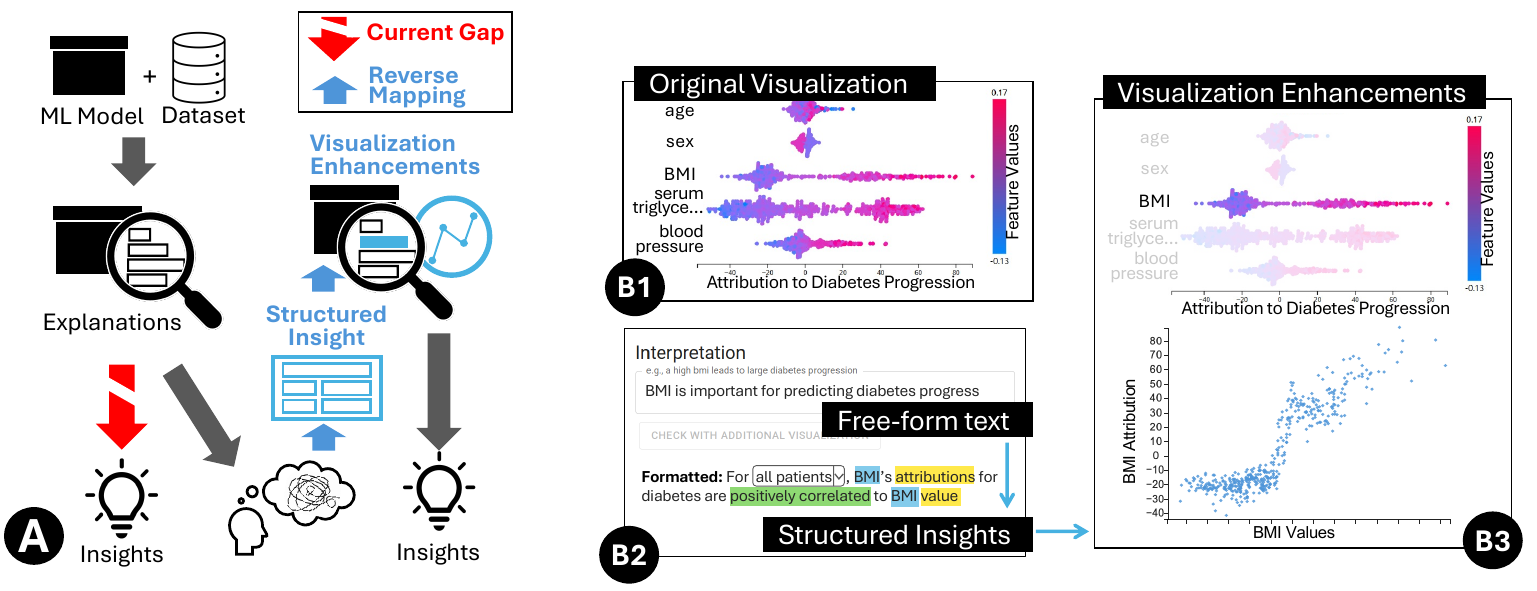}
  \vspace{-1em}
  \caption{
  \textbf{Reverse Mapping to enhance the interpretation of AI explanations.}
(A) Traditional XAI workflows rely on users to interpret visual explanations and derive insights, often leading to misinterpretation or over-trust. We propose a \textit{Reverse Mapping} paradigm that closes the loop by integrating user-generated insights back into the visual explanation process.
(B1) Original XAI visualizations show feature attributions but may not support nuanced interpretation.
(B2) Users express free-form insights which are automatically parsed into structured representations.
(B3) These structured insights are used to enhance the original visualizations with interaction annotations and additional visualizations.
  }
  \label{fig:teaser}
}

\abstract{
   \input{sections/0_abstract}
} % end of abstract


\begin{document}


\maketitle

\input{sections/1_Intro}
\input{sections/2_RelatedWork}

\input{sections/3_Method}
\input{sections/4_UseCases}
\input{sections/5_QualitativeFeedback}
\input{sections/6_Conclusion}

% The authors wish to thank A, B, and C. This work was supported in part by
% a grant from XYZ.}

%\bibliographystyle{abbrv}
\bibliographystyle{abbrv-doi}

\bibliography{template}
\end{document}

%% file: command.tex
\usepackage{enumitem}
\usepackage{flushend}

\newcommand{\ie}{\textit{i.e.}}
\newcommand{\eg}{\textit{e.g.}}
\newcommand{\etal}{\textit{et al.}}

\newcommand{\change}[1]{\textcolor{black}{#1}}

%% file: sections/0_abstract.tex
As AI systems become increasingly integrated into high-stakes domains, enabling users to accurately interpret model behavior is critical. 
While AI explanations can be provided, users often struggle to reason effectively with these explanations, limiting their ability to validate or learn from AI decisions. 
To address this gap, we introduce \textit{Reverse Mapping}, a novel approach that enhances visual explanations by incorporating user-derived insights back into the explanation workflow. 
Our system extracts structured insights from free-form user interpretations using a large language model and maps them back onto visual explanations through interactive annotations and coordinated multi-view visualizations. 
Inspired by the verification loop in the visualization knowledge generation model, this design aims to foster more deliberate, reflective interaction with AI explanations. 
We demonstrate our approach in a prototype system with two use cases and qualitative user feedback. 

%% file: sections/1_Intro.tex
\section{Introduction}
As AI becomes ubiquitous across domains, understanding AI is crucial for establishing appropriate trust, especially in high-stakes areas like healthcare~\cite{wang2022extending}. 
Beyond this, given the vast knowledge encoded within AI models, interpreting these systems holds the promise of uncovering insights beyond current human understanding.

The field of Explainable AI (XAI) has expanded rapidly in response to growing demands for AI transparency. Prior research has demonstrated the potential benefits of XAI, such as identifying subpopulations with shared mortality risks despite divergent feature values~\cite{lundberg2020local} and uncovering novel features associated with early stages of disease not previously noted by human clinicians~\cite{waldstein2020unbiased}. 
However, numerous user studies reveal that AI explanations often fall short, failing in tasks such as debugging AI models~\cite{adebayo2020debugging} and improving human decision-making~\cite{bansal2021does}.

XAI's failure to achieve its intended purposes has been attributed largely to users’ inability to properly interpret AI explanations~\cite{kaur2020interpreting, buccinca2021trust}.
% , relying on System 1 thinking (fast and intuitive) rather than System 2 thinking (deliberate and analytical)
Users often blindly trust AI explanations, failing to critically reason with the provided explanations.
Several studies have promoted more deliberate reasoning through interventions like cognitive forcing functions~\cite{buccinca2021trust}, incremental information disclosure~\cite{bo2024incremental}, and gamification~\cite{you2024gamifying}.
Comparatively little attention has been given to visualizations' role in facilitating AI understanding, despite studies showing that users perceive, interpret, and act upon AI-generated insights differently from the visual representations of explanations~\cite{wang2022extending, yang2020visual}.

\change{
To enable accurate XAI interpretation, this study proposes to adaptively enhance the XAI visualizations directly based on user-derived insights.
This reverse mapping (\ie, from insights to visualizations) inverts the typical flow, where users extract insights from visualizations (\cref{fig:teaser}A).
}
Specifically, the proposed method converts free-form, unstructured text into structured insights, which are then mapped to the visual explanation, including enhancements such as interactive annotations and coordinated multi-view visualizations. 
\change{
These enhancements help users better understand XAI and verify specific insight, echoing how the knowledge validation occurs through revisiting visualizations in the visualization knowledge generation model~\cite{sacha2014knowledge}.
}
% \qw{ADD more details about evaluation here later}
Main contributions of this paper are:
\begin{itemize}[leftmargin=*, nosep]
    \item The \textit{Reverse Mapping} paradigm, which enhances users’ interpretation of AI explanations by mapping user-derived insights back onto visual explanations.
    \item A prototype tool implementing the \textit{Reverse Mapping} paradigm through LLM-powered structured insight extraction, interactive annotations, and coordinated multi-view visualizations.
    \item Initial evidence that demonstrates the usability and effectiveness of \textit{Reverse Mapping} via use cases and qualitative user feedback.
\end{itemize}

%% file: sections/2_RelatedWork.tex
\section{Related Studies}

% \cite{lo2023change} investigate different methods for illustrating Poorly Constructed Visualization Designs.

% ProReveal~\cite{jo2019proreveal} provide a means of ensuring the correctness of the conclusion and understanding the reason when intermediate knowledge of visual analytics becomes invalid.  
% ``Value, Rank, Range, Comparative, Power Law, Normal, and Linear.''

% "The ten low-level tasks:
% • Retrieve Value
% • Filter
% • Compute Derived Value
% • Find Extremum
% • Sort
% • Determine Range
% • Characterize Distribution
% • Find Anomalies
% • Cluster
% • Correlate
% "

% low level task in visualization "Retrieve Value, Filter, Compute Derived Values, Find Extremum, Sort, Determine Range, Characterize Distribution, Find Anomalies, Cluster, and Correlate."
Visualizations have been widely established as an effective medium for communicating AI explanations~\cite{hohman2019gamut, yuan2020survey} due to their capacity to represent complex information in accessible formats. 
Popular XAI libraries such as SHAP~\cite{lundberg2020local} and DICE~\cite{mothilal2020explaining} have all incorporated visualization support as core components, enabling AI researchers and practitioners to generate static visual representations of computational explanations.
\change{Such tools have been incorporated into many interactive systems that help users understand AI explanations.} These visualization systems have demonstrated effectiveness across diverse tasks,  
including data augmentation and cleaning~\cite{chen2021oodanalyzer, yang2020drift}, model debugging~\cite{cao2021analyzing}, and model comparison and selection~\cite{wang2019atmseer, wang2019genealogy}. The interactive nature of these tools enhances users' abilities to explore and understand AI systems beyond the limitations of static visualizations.

As technical advances in XAI visualization have progressed, researchers have increasingly adopted user-centered perspectives, recognizing that technical sophistication alone is insufficient for effective explanations. 
Studies reveal concerning patterns where XAI visualizations can inadvertently encourage blind trust rather than critical analytical reasoning~\cite{kaur2020interpreting, bansal2021does}. 

This insight has prompted researchers to incorporate user needs and tasks into XAI visualization tool development ~\cite{ming2018rulematrix, wang2022extending}.
These studies contribute novel visualization designs and coordinated views to help domain users understand complicated data and generate meaningful insights.
For example, Wang \etal ~\cite{wang2022extending} found that visualizing paths for explaining AI explanations on drug repurposing significantly improved users' decision-making performance compared with other explanation visualizations.
However, these approaches have predominantly focused on designing optimized visualizations for predefined analytical tasks rather than using visualizations as dynamic mechanisms to actively mitigate incorrect interpretations of XAI outputs.

To address concerns about improper usage of explanation visualizations, researchers have proposed various intervention methods~\cite{buccinca2021trust, bo2024incremental, buccinca2024contrastive}. 
For example, Buccinca \etal ~\cite{buccinca2024contrastive} demonstrated that constrasting AI recommendations with predicted human choices significantly enhanced independent decision-making compared to unilateral explanations. While these interventions show promise, existing research has overlooked the potential for direct and dynamic modification of explanation visualizations to address misinterpretation. This gap represents our primary focus: exploring how adaptive visualization techniques can actively mitigate misunderstandings in explanations.

%% file: sections/3_Method.tex
\section{The Gaps between Explanation Visualizations and Insights}
The transition from visual representation to meaningful insight presents significant challenges in AI explanation interpretation, with three critical gaps.
First, \textbf{low visualization literacy \change{(Gap1)}} creates fundamental barriers when users fail to correctly identify and decode visual elements~\cite{kaur2020interpreting}. This manifests when users misattribute meaning to incorrect visual channels, \eg, attempting to extract feature values from axis position when this information is encoded through color. Such misinterpretations lead to fundamentally flawed conclusions about the underlying AI behavior.
Second, \textbf{ineffective visualization \change{(Gap2)}} selection often occurs when the chosen visual representation is not suited for extracting specific types of insights. Even with accurate interpretation of visual elements, the visualization format itself may make certain insights difficult or impossible to discern.
Third, \textbf{ambiguous, untestable insights \change{(Gap3)}} present a significant challenge to meaningful interpretation. Users may develop observations that lack sufficient specificity to be empirically verified or falsified, providing a limited actionable understanding of the AI system's behavior~\cite{brundage2020toward}.

% These gaps collectively contribute to the discrepancy between explanation visualization and effective insight generation, highlighting the need for approaches assisting XAI interpretation.

\section{Methods}

% To support the \textit{Reverse Mapping} paradigm, we developed a system that operationalizes this approach through a combination of structured insight extraction, visualization enhancement techniques, and an interactive user interface. 

% \subsection{Method Overview}

As illustrated in \autoref{fig:teaser}, the \textit{Reverse Mapping} paradigm is implemented through a two-step process.
% First, users observe explanation visualizations and document their observations in free-form text. These texts, often containing ambiguities or incomplete interpretations, are then converted into a structured insight representation, where uncertainties and incomplete understandings are highlighted for user verification and refinement.
% Second, these structured insights are mapped back to the original visualizations through interactive annotations or coordinated multi-views, which enables users to validate their understanding and identify misinterpretations.
First, users review explanation visualizations and describe their observations in free-form text. These often-ambiguous interpretations are then translated into structured insights, highlighting uncertainties for user verification. Second, structured insights are linked back to the visualizations via interactive annotations, highlights, and/or coordinated views, helping users validate or refine their understanding. 

% While there exists a wide array of XAI methods, this study strategically focuses on global explanations and attribution-based explanations to illustrate the \textit{Reverse Mapping} paradigm and implement the prototype tool. 
% This focus is strategic for three reasons.
% First, global explanations present greater interpretive challenges than local explanations, as they characterize AI behavior across the entire input space rather than single predictions.
% Second, global explanations reveal diverse insights that cannot be observed through individual predictions, enabling richer visualization and analytics opportunities.
% Third, attribution-based explanations, which quantify each feature's contribution to outputs, offer model-agnostic applicability and are widely used.

This study focuses on global and attribution-based explanations to demonstrate Reverse Mapping. Global explanations, which describe model behavior across the input space, pose greater interpretive challenges and reveal richer insights than local ones. Attribution methods, like SHAP~\cite{lundberg2020local}, quantify feature contributions, are model-agnostic, and widely adopted.
\change{The visualization types in this study are derived from SHAP, with interactive versions implemented based on originally static designs. 
While representing only a subset of XAI visualizations, these visualizations are widely-used representative examples and thus provide a suitable testbed for instantiating \textit{ReverseMapping} and evaluating the prototype tool.
}

Our prototype uses SHAP with grammar-based visualizations (\eg, Vega-Lite~\cite{satyanarayan2016vega}, Gosling~\cite{lyi2021gosling}), but the method is compatible with any attribution-based explanation technique.
\change{Implementation details are available in the supplementary materials.}
% The source code is available at \url{URL_hidden_for_anonymous_review}.
% In our prototype implementation, we used SHAP explanations~\cite{lundberGap2020local} and assumed the original visualizations come with grammar-based visualization specifications (\eg, Vega-Lite~\cite{satyanarayan2016vega}, Gosling~\cite{lyi2021gosling}) that explicitly specify the mapping from data values into visual marks and channels. 
% However, the proposed method is model-agnostic and can be applied to any attribution-based AI explanation method.

\subsection{The Space of Structured Insights}
A structure space of XAI insights serves as a critical mediation for bridging the free-form user observations to the enhancements of visualizations.
While prior studies have summarized the key XAI elements and intelligibility queries~\cite{liao2020questioning, wang2019designing}, there remains a significant gap about  the types of insights users infer from explanations.
% In this subsection, we will introduce this structured space using the explanation of a model that predicts patients' diabetes progression based on input features such as age, sex, BMI, and blood pressure.

To structure the insight space, we examine the data format of attribution-based global explanations. 
Typically, this involves tabular data where each row represents a data point, encompassing feature values, predictions, and the corresponding feature attributions. While each row provides a local explanation for an individual prediction, aggregating these local explanations across numerous instances forms global explanations of the model's behavior, as demonstrated by Lundberg \etal.~\cite{lundberg2020local}.
We contextualize the data format of attribution-based global explanations using task abstraction in visual analytics~\cite{munzner2014visualization}. 
Based on this, we identify three primary insight types, all centered around feature attributions (the key component of explanations): 
\begin{itemize}[leftmargin=*, nosep]
    \item \textbf{Read.} Extract values related to the attribution of a feature to the prediction (\eg, average attribution, variations of attribution, number of data points with positive attribution). For example, identifying that, for more than 65\% of patients, \textit{BMI} has a positive attribution to the predicted \textit{diabetes progression}.
    \item \textbf{Correlation.} Building upon Read, identify relationships between feature values and their corresponding attribution-related values. For example, discovering that as \textit{age} increases, the attribution of \textit{age} to \textit{diabetes progression} also tends to increase. 
    \item \textbf{Comparison.} Building upon Read, compare attribution-related values of different features for a given prediction. For example, finding that \textit{BMI} on average has a larger positive attribution to the predicted \textit{diabetes progression} than \textit{blood pressure}.
\end{itemize}
Insights can also be conditioned on feature value ranges: For example, ``... when age is above 65" or ``... when BMI is below 0."

\begin{figure}
    \centering
    \includegraphics[width=\linewidth]{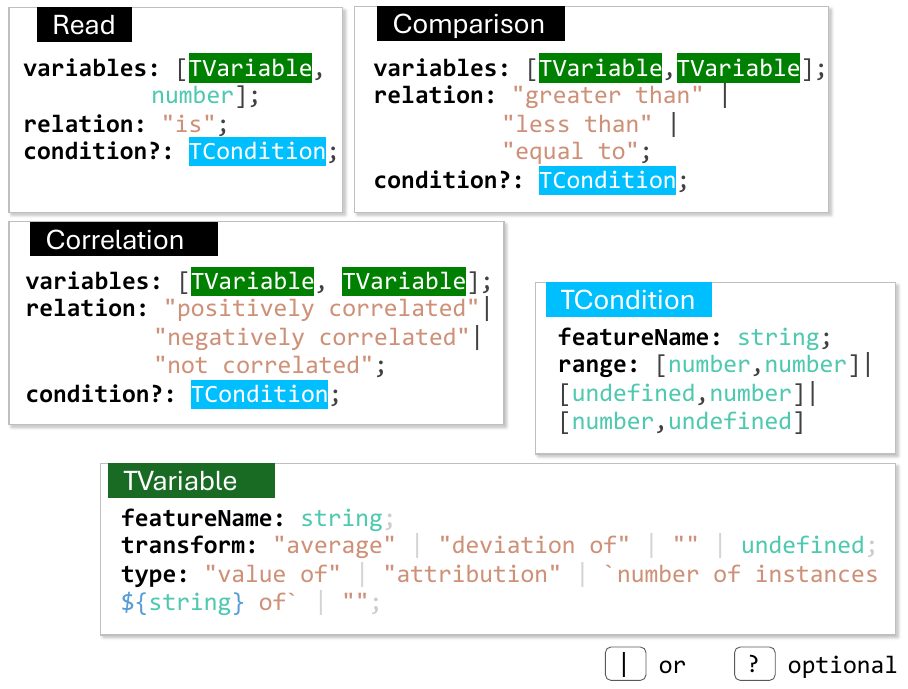}
    \vspace{-2em}
    \caption{\textbf{Schema of the Structured Insight Format.} The three insight types, \ie, Read, Comparison, and Correlation, are specified by variables, relations, and optional conditions. Each variable (\texttt{TVariable}) and condition (\texttt{TCondition}) follows a consistent format to support structured parsing. The schema follows the type definition syntax in TypeScript.}
    \label{fig:json}
\end{figure}

\subsection{Free-Form Text to Structured Insights}

The structured insight space provides a foundation for converting user-provided free-form text (Gap3) into a consistent, unified format. This conversion uses OpenAI’s GPT-4o model through prompt engineering techniques, including few-shot and chain-of-thought prompting.
The LLM first classifies the insight type \change{(read, correlation, comparison)} and next determines whether there is a feature range condition. Definitions and examples of each insight type are embedded within prompts to guide classification.
Second, the system transforms the free-form text into a predefined JSON template corresponding to the classified insight type, as illustrated in \autoref{fig:json}. 
This template-filling step also identifies potential incompleteness in the user input, serving as an initial mechanism for users to reflect on and refine their insights. 

\begin{figure}
    \centering
    \includegraphics[width=\linewidth]{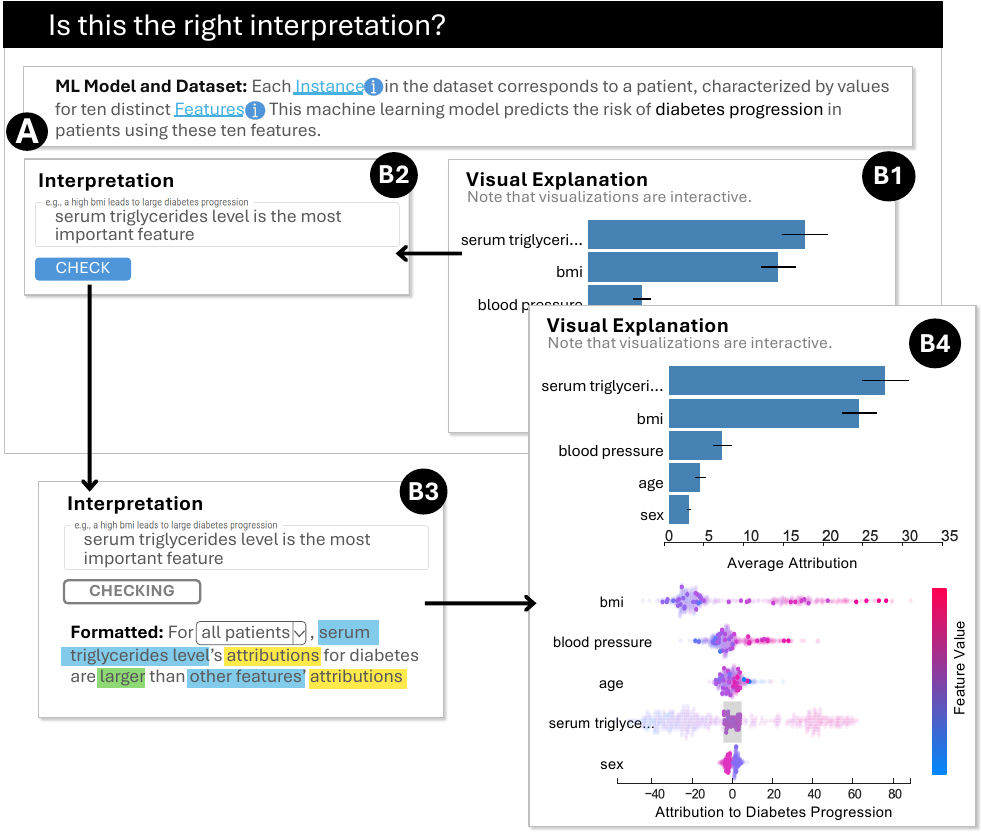}
    \vspace{-2em}
    \caption{The \textbf{Interface} includes (A) a Model Card summarizing the AI model, \change{(B2 \& B3)} an Insight Card for entering and structuring user interpretations, and \change{(B1 \& B4)} an Explanation Visualizer showing the original and enhanced visual explanations. }
    \label{fig:interface}
\end{figure}

To enhance readability and user interaction, a rule-based approach converts the JSON structure back into natural language sentences with interactive elements. 
As shown in \change{\autoref{fig:interface}B3}, key components such as feature names, attribution, insight types, and conditions are highlighted, while missing or ambiguous values are presented as dropdown menus or input fields, allowing users to easily identify and complete incomplete information.

\subsection{Reverse Mapping from Insight to Visualization}
Based on the structured insight, we consider two main types of enhancement to the original visualizations: i) annotations upon the original visualization, which highlight the corresponding graphical elements related to the insight (Gap1); and ii) an additional visualization that coordinated with the original visualization to better illustrate the given insight (Gap2).

To generate the annotations, we begin by aligning the structured insight with the underlying visualization specifications to identify the relevant visual marks (\eg, points, bars) and encoding channels (\eg, color, x-position).
\change{For instance, if an insight relates to \textit{age}, the associated visual marks and channels can be identified by examining their associations with the feature \textit{age} in visualization specifications.}
Once identified, these marks and the explanatory components of channels, such as legends for color encodings or axes for positional encodings, are visually emphasized by dimming unrelated elements through reduced opacity.
When the structured insight includes constant values (\eg, when age is larger than 65), the system overlays dashed reference lines at the corresponding positions to aid interpretation.

The generation of additional visualizations draws inspiration from knowledge-based visualization recommendation systems~\cite{pandey2022genorec}. 
Rather than relying solely on LLMs to suggest visualizations based on insights, which can lead to preferences that diverge from empirical findings on human interpretability~\cite{wang2024dracogpt}, \change{we incorporate relevant design guidelines summarized by previous studies~\cite{pandey2022genorec, wang2024dracogpt} into the prompting to guide visualization selection. 
These guidelines specify which visual encodings are most effective for different analytical tasks and insight categories, steering visualization generation toward representations aligned with established visualization design best practices. 
For example, for correlation-related insights, the prompt encourages the use of scatter plots rather than bar charts, as scatter plots are empirically shown to better convey relationships between continuous variables.
}

\subsection{Interface}

As shown in \autoref{fig:interface}, the interface is composed of several key components: a \textit{Model Card} (A), which presents basic information about the AI model such as input and output data; an \textit{Insight Card} \change{(B2, B3)}, where users can input free-form text to extract structured insights; and an \textit{Explanation Visualizer} \change{(B1, B4)}, which displays both the original visualization and enhancements generated through \textit{Reverse Mapping}. 
% \change{Visualization types and styles are drawn from SHAP \cite{lundberg2020local}.}
Users begin by reviewing the Model Card and the initial visualization \change{(B1)}. They then enter their observations into the text field on the Insight Card \change{(B2)}. Upon clicking the ``Check'' button, the system automatically processes the input to generate a structured insight. Keywords and potentially missing values are highlighted for clarity. The corresponding insights are then reflected as interactive annotations on the original visualization \change{(B3)}. If the initial visualization is not well-suited to represent the given insight, an additional visualization is generated and coordinated with the original visualization to form a multi-view visualization \change{(B4)}.

%% file: sections/4_UseCases.tex
\section{Use Cases}

% OLD WORK
% In order to evaluate this system, we developed a user study. We crafted a series of insight statements, half true and half false. These statements ranged in variety, covering all the insight categories described previously. For each statement, we provided a baseline visualization: One that provided some information, but was not the optimal visualization for that statement. 

% For each statement, we then had two conditions: Random and Optimal. In the random case, we used a random number generator to determine a visualization (non-optimal) that would be shown alongside the baseline one. 

% In the optimal case, we used our reverse mapping rules to determine the optimal visualization for that particular statement, and then displayed that visualization alongside the baseline one. 

% Users were shown each statement (with just the baseline visualization) and were asked to determine whether it was true or false. They were then asked to rate their confidence. Then, users were shown a second visualization (either random or optimal). THen they were asked to evalate the statement's veracity (and their confidence) once again. 

% We present two different use cases with examples for the reverse mapping tool. In the first use case, the user makes a correct initial insight, whereas in the second use case, the user makes an incorrect initial insight.

This section presents two use cases that demonstrate the utility of \textit{Reverse Mapping} in both validating and rejecting user interpretation of AI explanations.
We use a machine learning model that predicts patients’ diabetes progression based on ten input features, including age, sex, BMI, and blood pressure.

\begin{figure}
    \centering
    \includegraphics[width=\linewidth]{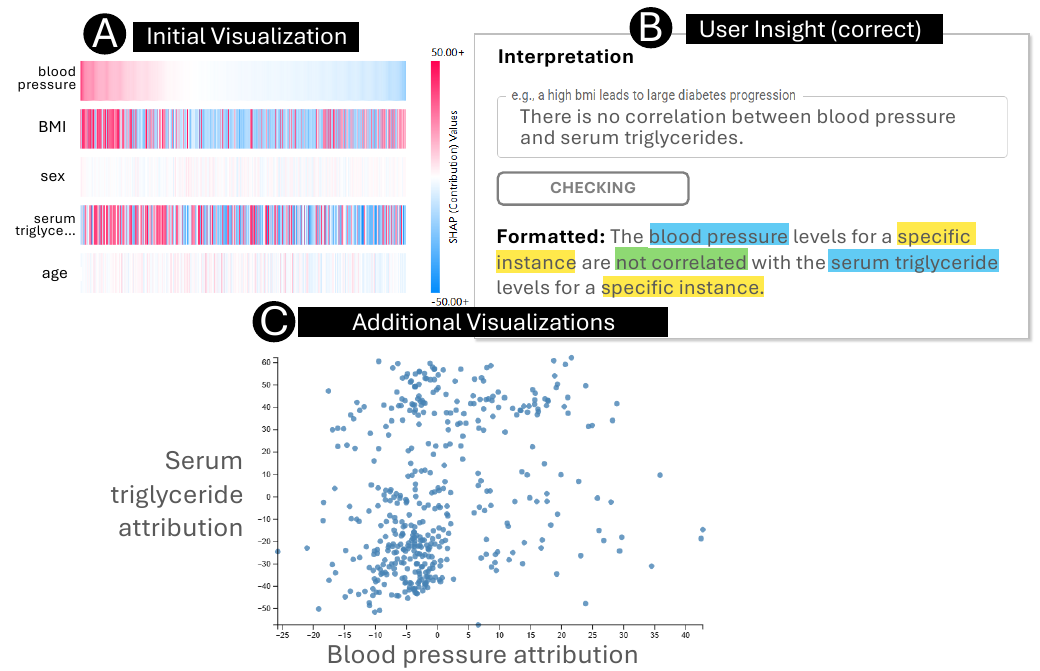}
    \vspace{-2em}
    \caption{\textbf{Case 1: Verify User Insight.} 
    The \textit{Reverse Mapping} enhances the original heatmap plot with a scatter plot, which more effectively illustrates the lack of correlation between the two variables and reinforce the insight.}
    
    %An additional visualization is provided to help further verify the insight generated with the original heatmap visualization.
    
    % An example use case where a user is initially provided with a heatmap visualization. In this use case, the user makes a correct insight, and the additional visualizations provided from \textit{Reverse Mapping} reinforce this. Additional annotations have been added to direct attention towards the specific parts of the additional visualizations that verify the insight.
    \label{fig:correct-case}
\end{figure}

\autoref{fig:correct-case} shows a use case where \textit{Reverse Mapping} helps reinforce an insight. 
The user first sees a heatmap (A), where cell color represents feature attributions, with rows for features and columns for instances.
Noticing that blood pressure (top row) and serum triglycerides (fourth row) do not share similar color gradient, the user believes ``there is no correlation between blood pressure attributions and serum triglycerides attributions.'' Upon clicking the ``Check'' button, the system reformulates this observation and generates a scatter plot (C) with blood pressure attribution on the x-axis and serum triglyceride attribution on the y-axis. The resulting plot shows no discernible pattern, reinforcing the user’s original insight.

% An example of the first use case is in \autoref{fig:correct-case}, where a user is initially presented with a heatmap visualization (A) showing the relationships between different features across specific instances. After reviewing the visualization, the user notices the instances for blood pressure and serum triglycerides do not share similar color intensities, leading to the claim that ``There is no correlation between blood pressure and serum triglycerides'' (B).  The tool then transforms the insight into a testable statement, such as, ``The blood pressure levels for a specific instance are not correlated with the serum triglyceride levels for a specific instance.'' Our tool explicitly highlights keywords in the testable statement, like the feature names, correlation type, and testable metrics. Then, the tool generates a second visualization to test their claim. One possibility is a scatter plot where the blood pressure attribution is the x-axis, and the serum triglyceride attribution is the y-axis (C). In this case, the user can quickly see the lack of correlation between the two features, reinforcing their original insight.

\begin{figure}
    \centering
    \includegraphics[width=\linewidth]{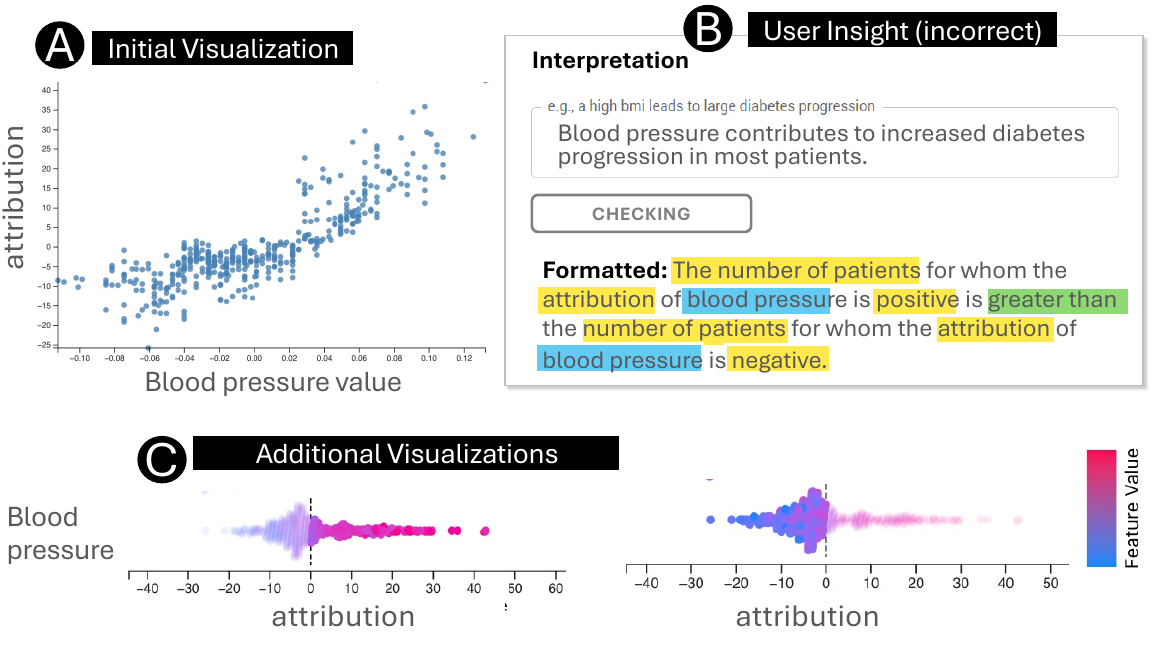}
    \vspace{-2em}
    \caption{\textbf{Case 2: Reject User Insight.} 
    The \textit{Reverse Mapping} enhances the original scatter plot with a beeswarm plot, which more effectively illustrates the distribution of data points and thus reveals the inaccuracy of the input insight about the number of patients.
    % An example use case where a user is initially provided with a scatter visualization. In this use case, the user makes an incorrect insight, and the additional visualizations provided from \textit{Reverse Mapping} refute this. Additional annotations have been added to direct attention towards the specific parts of the additional visualizations that refute the insight.
    }
    \label{fig:incorrect-case}
\end{figure}

\autoref{fig:incorrect-case} shows another use case where \textit{Reverse Mapping} helps identify an incorrect insight. The user first examines a scatter plot (A) showing the relationship between blood pressure values and their attributions to diabetes progression. By visually estimating the data distribution, the user perceives more points with positive attribution and states: ``Blood pressure contributes to increased diabetes progression in most patients'' (B). Upon clicking ``Check'', the system reformulates this vague statement into a testable comparison: ``The number of patients with positive attribution for blood pressure is greater than the number with negative attribution.'' 
To support verification, the system generates a dual beeswarm plot (C), which more effectively conveys data point quantities. 
The left side's larger shape area illustrates the original insight's inaccuracy.

% Next, an example of the second use case is in \autoref{fig:incorrect-case}, where a user is initially presented with a scatter plot showing the distribution of blood pressure instances. The user may initially see the visualization and claim ``Blood pressure has more positive instances than negative instances.'' Once again, the tool restructures the insight as a testable statement: ``The number of instances above 0 for blood pressure is greater than the number of instances below 0 for blood pressure,'' where key aspects of the testable statement are highlighted. The tool then generates an additional visualization to test their claim. One possibility is a dual beeswarm plot with an annotation showing only the negative instances and another plot with an annotation showing only the positive instances. From the ``Selected Points'' annotation, the user can clearly see there are more negative instances of blood pressure than positive, contradicting their original insight.

% By utilizing this process of continuous feedback, the tool forces users to practice reasoning skills through the validation of their own interpretations.

%% file: sections/5_QualitativeFeedback.tex
\section{\change{Preliminary} User Feedback}
To evaluate the utility of the \textit{Reverse Mapping}, we conducted an \change{tentative} online user study with 25 participants \change{recruited from Prolific} (13 female, 12 male, mean age = 37.48, SD = 9.76), who completed the study independently without supervision.
\change{These participants had a mean self-reported machine learning experience of 3.16/6, with a mode of 1.}
% The goal was to evaluate the utility of the \textit{Reverse Mapping} mechanism in supporting the interpretation of AI explanations. 
 % The study aimed to investigate two main research questions: (1) To what extent can users interpret insights from ? and (2) How do visualization enhancements generated through Reverse Mapping influence users' understanding and error detection?
% To ensure a controlled setting, participants did not type free-form input from scratch but were offered a set of 12 visualizations and corresponding insights to reduce entry barriers. \change{GPT-4o was not used for reverse-mapping; each insight's JSON was hand-written and verified.}

Among the participants, 44\% reported being able to interpret the insights based on the initial visualizations alone. 
However, 84\% indicated that the visualization enhancements generated through \textit{Reverse Mapping} were ``very helpful'' in deepening their understanding.
% while an additional 8\% found them ``occasionally helpful''.
When the initial statement was incorrect and participants did not recognize the error with the original visualization, 56\% of the time they were able to identify the mistake after viewing the enhanced visualizations from \textit{Reverse Mapping}.

% 56\% of participants also indicated that only having one visualization was insufficient to understand the provided insight.

% We additionally conducted interviews with three users who did not take the user study. 
We additionally conducted interviews with three participants who completed the same study protocol under supervision, allowing them to ask questions and think aloud throughout the process.
% All three were male and aged 20, 21, and 58. 
% Their feedback can be divided into three categories: ideal usage, benefits, and shortcomings.
Two users indicated their ideal usage was applying the tool on their own datasets to validate insights they may already have. Additionally, one user noted they would utilize the tool for educational purposes, such as understanding how some visual representations can be misleading without more context.
Several benefits of the tool were identified. 
Two users stated that the ability to interact with explanations, inputting insights and seeing corresponding visualization updates, is more enjoyable compared to passively reading static outputs.
% One user stated that typing their insights acted as a forcing function for critical thinking about the dataset.
One user stated that typing their insights acted as trigger for him to critically think about the dataset.
% All three users noted that visual encodings were more memorable than text alone, supporting a deeper retention of insights. 
Furthermore, all three users agreed the tool encouraged careful thinking and analysis about the explanations, rather than mindlessly reliance.
% a specific issue in the field of XAI \cite{buccinca2021trust}. 

At the same time, users also pointed out several limitations and offered valuable suggestions for improvement.
They noted that the tool occasionally generated visualization enhancements whose connection to 
the underlying insights was unclear. For example, one user received a heatmap visualization when examining age attribution in a subgroup, leading to confusion about how to identify such subgroups in visualization. More detailed guidance on interpreting these enhancements would be helpful.
All three participants suggested incorporating a feedback loop to score the visualizations quality to refine and personalize the \textit{Reverse Mapping} process.
Additionally, they stated that effective usage requires basic understanding with visualization and XAI, which may limit its accessibility to a broader audience.

% At times, the tool would create visualizations that two of the users did not find helpful. Furthermore, one of the users' insights required more advanced visualizations, which the user found difficult to interpret. 
% Finally, one user expressed concern some users may inaccurately assume a correlation with causation from insights derived using the tool.

% Overall, the demographic diversity of our participants allowed us to see a wide range of perspectives and opinions on our reverse mapping mechanism. Users consistently found benefit in having a second visualization catered to helping them understand different insights, though its effectiveness was dependent both on the mapping used and the users' prior familiarity with data visualization. 

%% file: sections/6_Conclusion.tex
\section{Conclusion and Future Work}

% Future work:
% 1. user study to quantify the benefits of Reverse Mapping
% 2. More detailed and tailored assistance in reading the enhanced visualizations, such as textual explanations why an visualizations might reject the insights
% 3. Current version mainly focus on the visualization types supported in SHAP for the prototype implementation, a comprehensive understanding of the space of XAI visualizations

In this work, we introduced \textit{Reverse Mapping}, a novel paradigm that enhances user interpretation of AI explanations by integrating user-derived insights back into the explanation process. 
By structuring free-form user observations and mapping them to annotations and coordinated visualizations, this approach aims to bridge the gap between explanations and accurate and testable insight. 

% \change{While we have limited analytical power due to the small sample size and the qualitative nature of the user task, we believe that this approach offers an excellent starting point.}
% We implemented this paradigm in a prototype system and demonstrated its feasibility through illustrative examples and qualitative user feedback.

% Future work will focus on several directions. First, we aim to conduct empirical user studies \change{with larger study populations} to quantitatively evaluate the effectiveness of \textit{Reverse Mapping}. 
% Second, we plan to develop more nuanced and context-aware text guidance to assist users in interpreting the enhanced visualizations \change{and refine their hypotheses}. 
% Finally, we aim to conduct a comprehensive exploration of the design space of XAI visualizations to extend the generalizability and applicability of the proposed approach across different explanation methods and domains.
% \change{While our initial study offers promising results, its analytical power is limited due to the small sample size and the qualitative nature of the user task, which highlights the need for further validation.}

\change{Despite promising initial results,
this study has several limitations that open up directions for future research. 
First, the current evaluation is preliminary and qualitative. We plan to conduct empirical user studies with larger study populations to quantitatively evaluate the effectiveness of \textit{Reverse Mapping}. 
Second, using LLM in the \textit{ReverseMapping} requires further investigation and validation. The current usage of LLM in formatting user inputs can introduce unexpected errors. At the same time, LLMs also show promise in further supporting interpretation by generating textual explanations and counter-arguments. 
Finally, we aim to conduct a comprehensive exploration of the design space of XAI visualizations to extend the generalizability and applicability of the proposed approach across different explanation methods and visualizations.
}

%% file: template.bbl
\begin{thebibliography}{10}

\bibitem{adebayo2020debugging}
J.~Adebayo, M.~Muelly, I.~Liccardi, and B.~Kim.
\newblock Debugging tests for model explanations.
\newblock In {\em Proceedings of the 34th International Conference on Neural Information Processing Systems}. Curran Associates Inc., Red Hook, NY, USA, 2020.

\bibitem{bansal2021does}
G.~Bansal, T.~Wu, J.~Zhou, R.~Fok, B.~Nushi, E.~Kamar, M.~T. Ribeiro, and D.~Weld.
\newblock Does the whole exceed its parts? the effect of ai explanations on complementary team performance.
\newblock In {\em Proceedings of the 2021 CHI Conference on Human Factors in Computing Systems}. Association for Computing Machinery, New York, NY, USA, 2021. doi: {{%
10\hspace{.1pt}\discretionary{.}{%
}{.}\hspace{.4pt}1145\discretionary{/}{%
}{/}3411764\hspace{.1pt}\discretionary{.}{%
}{.}\hspace{.4pt}3445717}}


\bibitem{bo2024incremental}
J.~Y. Bo, P.~Hao, and B.~Y. Lim.
\newblock Incremental xai: Memorable understanding of ai with incremental explanations.
\newblock In {\em Proceedings of the 2024 CHI Conference on Human Factors in Computing Systems}. Association for Computing Machinery, New York, NY, USA, 2024. doi: {{%
10\hspace{.1pt}\discretionary{.}{%
}{.}\hspace{.4pt}1145\discretionary{/}{%
}{/}3613904\hspace{.1pt}\discretionary{.}{%
}{.}\hspace{.4pt}3642689}}


\bibitem{brundage2020toward}
M.~Brundage, S.~Avin, J.~Wang, H.~Belfield, G.~Krueger, G.~Hadfield, H.~Khlaaf, J.~Yang, H.~Toner, R.~Fong, et~al.
\newblock Toward trustworthy ai development: Mechanisms for supporting verifiable claims, 2020.

\bibitem{buccinca2021trust}
Z.~Bu\c{c}inca, M.~B. Malaya, and K.~Z. Gajos.
\newblock To trust or to think: Cognitive forcing functions can reduce overreliance on ai in ai-assisted decision-making.
\newblock {\em Proc. ACM Hum.-Comput. Interact.}, 5(CSCW1), Apr. 2021. doi: {{%
10\hspace{.1pt}\discretionary{.}{%
}{.}\hspace{.4pt}1145\discretionary{/}{%
}{/}3449287}}


\bibitem{buccinca2024contrastive}
Z.~Bu\c{c}inca, S.~Swaroop, A.~E. Paluch, F.~Doshi-Velez, and K.~Z. Gajos.
\newblock Contrastive explanations that anticipate human misconceptions can improve human decision-making skills.
\newblock In {\em Proceedings of the 2025 CHI Conference on Human Factors in Computing Systems}. Association for Computing Machinery, New York, NY, USA, 2025. doi: {{%
10\hspace{.1pt}\discretionary{.}{%
}{.}\hspace{.4pt}1145\discretionary{/}{%
}{/}3706598\hspace{.1pt}\discretionary{.}{%
}{.}\hspace{.4pt}3713229}}


\bibitem{cao2021analyzing}
K.~Cao, M.~Liu, H.~Su, J.~Wu, J.~Zhu, and S.~Liu.
\newblock Analyzing the noise robustness of deep neural networks.
\newblock {\em IEEE Transactions on Visualization and Computer Graphics}, 27(7):3289--3304, 2021. doi: {{%
10\hspace{.1pt}\discretionary{.}{%
}{.}\hspace{.4pt}1109\discretionary{/}{%
}{/}TVCG\hspace{.1pt}\discretionary{.}{%
}{.}\hspace{.4pt}2020\hspace{.1pt}\discretionary{.}{%
}{.}\hspace{.4pt}2969185}}


\bibitem{chen2021oodanalyzer}
C.~Chen, J.~Yuan, Y.~Lu, Y.~Liu, H.~Su, S.~Yuan, and S.~Liu.
\newblock Oodanalyzer: Interactive analysis of out-of-distribution samples.
\newblock {\em IEEE Transactions on Visualization and Computer Graphics}, 27(7):3335--3349, 2021. doi: {{%
10\hspace{.1pt}\discretionary{.}{%
}{.}\hspace{.4pt}1109\discretionary{/}{%
}{/}TVCG\hspace{.1pt}\discretionary{.}{%
}{.}\hspace{.4pt}2020\hspace{.1pt}\discretionary{.}{%
}{.}\hspace{.4pt}2973258}}


\bibitem{hohman2019gamut}
F.~Hohman, A.~Head, R.~Caruana, R.~DeLine, and S.~M. Drucker.
\newblock Gamut: A design probe to understand how data scientists understand machine learning models.
\newblock In {\em Proceedings of the 2019 CHI Conference on Human Factors in Computing Systems}, p. 1–13. Association for Computing Machinery, New York, NY, USA, 2019. doi: {{%
10\hspace{.1pt}\discretionary{.}{%
}{.}\hspace{.4pt}1145\discretionary{/}{%
}{/}3290605\hspace{.1pt}\discretionary{.}{%
}{.}\hspace{.4pt}3300809}}


\bibitem{kaur2020interpreting}
H.~Kaur, H.~Nori, S.~Jenkins, R.~Caruana, H.~Wallach, and J.~Wortman~Vaughan.
\newblock Interpreting interpretability: Understanding data scientists' use of interpretability tools for machine learning.
\newblock In {\em Proceedings of the 2020 CHI Conference on Human Factors in Computing Systems}, CHI '20, p. 1–14. Association for Computing Machinery, New York, NY, USA, 2020. doi: {{%
10\hspace{.1pt}\discretionary{.}{%
}{.}\hspace{.4pt}1145\discretionary{/}{%
}{/}3313831\hspace{.1pt}\discretionary{.}{%
}{.}\hspace{.4pt}3376219}}


\bibitem{liao2020questioning}
Q.~V. Liao, D.~Gruen, and S.~Miller.
\newblock Questioning the ai: Informing design practices for explainable ai user experiences.
\newblock In {\em Proceedings of the 2020 CHI Conference on Human Factors in Computing Systems}, p. 1–15. Association for Computing Machinery, New York, NY, USA, 2020. doi: {{%
10\hspace{.1pt}\discretionary{.}{%
}{.}\hspace{.4pt}1145\discretionary{/}{%
}{/}3313831\hspace{.1pt}\discretionary{.}{%
}{.}\hspace{.4pt}3376590}}


\bibitem{lundberg2020local}
S.~Lundberg, G.~Erion, H.~Chen, A.~DeGrave, J.~Prutkin, B.~Nair, R.~Katz, J.~Himmelfarb, N.~Bansal, and S.-I. Lee.
\newblock From local explanations to global understanding with explainable ai for trees.
\newblock {\em Nature Machine Intelligence}, 2, 01 2020. doi: {{%
10\hspace{.1pt}\discretionary{.}{%
}{.}\hspace{.4pt}1038\discretionary{/}{%
}{/}s42256\discretionary{%
}{-}{-}019\discretionary{%
}{-}{-}0138\discretionary{%
}{-}{-}9}}


\bibitem{lyi2021gosling}
S.~L'Yi, Q.~Wang, F.~Lekschas, and N.~Gehlenborg.
\newblock Gosling: A grammar-based toolkit for scalable and interactive genomics data visualization.
\newblock {\em IEEE Transactions on Visualization and Computer Graphics}, 28(1):140--150, 2022. doi: {{%
10\hspace{.1pt}\discretionary{.}{%
}{.}\hspace{.4pt}1109\discretionary{/}{%
}{/}TVCG\hspace{.1pt}\discretionary{.}{%
}{.}\hspace{.4pt}2021\hspace{.1pt}\discretionary{.}{%
}{.}\hspace{.4pt}3114876}}


\bibitem{ming2018rulematrix}
Y.~Ming, H.~Qu, and E.~Bertini.
\newblock Rulematrix: Visualizing and understanding classifiers with rules.
\newblock {\em IEEE Transactions on Visualization and Computer Graphics}, 25(1):342--352, 2019. doi: {{%
10\hspace{.1pt}\discretionary{.}{%
}{.}\hspace{.4pt}1109\discretionary{/}{%
}{/}TVCG\hspace{.1pt}\discretionary{.}{%
}{.}\hspace{.4pt}2018\hspace{.1pt}\discretionary{.}{%
}{.}\hspace{.4pt}2864812}}


\bibitem{mothilal2020explaining}
R.~K. Mothilal, A.~Sharma, and C.~Tan.
\newblock Explaining machine learning classifiers through diverse counterfactual explanations.
\newblock In {\em Proceedings of the 2020 Conference on Fairness, Accountability, and Transparency}, p. 607–617. Association for Computing Machinery, New York, NY, USA, 2020. doi: {{%
10\hspace{.1pt}\discretionary{.}{%
}{.}\hspace{.4pt}1145\discretionary{/}{%
}{/}3351095\hspace{.1pt}\discretionary{.}{%
}{.}\hspace{.4pt}3372850}}


\bibitem{munzner2014visualization}
T.~Munzner.
\newblock {\em Visualization analysis and design}.
\newblock CRC press, 2014.

\bibitem{pandey2022genorec}
A.~Pandey, S.~L'Yi, Q.~Wang, M.~A. Borkin, and N.~Gehlenborg.
\newblock Genorec: A recommendation system for interactive genomics data visualization.
\newblock {\em IEEE Transactions on Visualization and Computer Graphics}, 29(1):570--580, 2023. doi: {{%
10\hspace{.1pt}\discretionary{.}{%
}{.}\hspace{.4pt}1109\discretionary{/}{%
}{/}TVCG\hspace{.1pt}\discretionary{.}{%
}{.}\hspace{.4pt}2022\hspace{.1pt}\discretionary{.}{%
}{.}\hspace{.4pt}3209407}}


\bibitem{sacha2014knowledge}
D.~Sacha, A.~Stoffel, F.~Stoffel, B.~C. Kwon, G.~Ellis, and D.~A. Keim.
\newblock Knowledge generation model for visual analytics.
\newblock {\em IEEE Transactions on Visualization and Computer Graphics}, 20(12):1604--1613, 2014. doi: {{%
10\hspace{.1pt}\discretionary{.}{%
}{.}\hspace{.4pt}1109\discretionary{/}{%
}{/}TVCG\hspace{.1pt}\discretionary{.}{%
}{.}\hspace{.4pt}2014\hspace{.1pt}\discretionary{.}{%
}{.}\hspace{.4pt}2346481}}


\bibitem{satyanarayan2016vega}
A.~Satyanarayan, D.~Moritz, K.~Wongsuphasawat, and J.~Heer.
\newblock Vega-lite: A grammar of interactive graphics.
\newblock {\em IEEE Transactions on Visualization and Computer Graphics}, 23(1):341--350, 2017. doi: {{%
10\hspace{.1pt}\discretionary{.}{%
}{.}\hspace{.4pt}1109\discretionary{/}{%
}{/}TVCG\hspace{.1pt}\discretionary{.}{%
}{.}\hspace{.4pt}2016\hspace{.1pt}\discretionary{.}{%
}{.}\hspace{.4pt}2599030}}


\bibitem{waldstein2020unbiased}
S.~Waldstein, P.~Seeböck, R.~Donner, A.~Sadeghipour, H.~Bogunović, A.~Osborne, and U.~Schmidt-Erfurth.
\newblock Unbiased identification of novel subclinical imaging biomarkers using unsupervised deep learning.
\newblock {\em Scientific Reports}, 10:12954, 07 2020. doi: {{%
10\hspace{.1pt}\discretionary{.}{%
}{.}\hspace{.4pt}1038\discretionary{/}{%
}{/}s41598\discretionary{%
}{-}{-}020\discretionary{%
}{-}{-}69814\discretionary{%
}{-}{-}1}}


\bibitem{wang2019designing}
D.~Wang, Q.~Yang, A.~Abdul, and B.~Y. Lim.
\newblock Designing theory-driven user-centric explainable ai.
\newblock In {\em Proceedings of the 2019 CHI Conference on Human Factors in Computing Systems}, p. 1–15. Association for Computing Machinery, New York, NY, USA, 2019. doi: {{%
10\hspace{.1pt}\discretionary{.}{%
}{.}\hspace{.4pt}1145\discretionary{/}{%
}{/}3290605\hspace{.1pt}\discretionary{.}{%
}{.}\hspace{.4pt}3300831}}


\bibitem{wang2024dracogpt}
H.~W. Wang, M.~Gordon, L.~Battle, and J.~Heer.
\newblock Dracogpt: Extracting visualization design preferences from large language models.
\newblock {\em IEEE Transactions on Visualization and Computer Graphics}, 31(1):710–720, Jan. 2025. doi: {{%
10\hspace{.1pt}\discretionary{.}{%
}{.}\hspace{.4pt}1109\discretionary{/}{%
}{/}TVCG\hspace{.1pt}\discretionary{.}{%
}{.}\hspace{.4pt}2024\hspace{.1pt}\discretionary{.}{%
}{.}\hspace{.4pt}3456350}}


\bibitem{wang2022extending}
Q.~Wang, K.~Huang, P.~Chandak, M.~Zitnik, and N.~Gehlenborg.
\newblock Extending the nested model for user-centric xai: A design study on gnn-based drug repurposing.
\newblock {\em IEEE Transactions on Visualization and Computer Graphics}, 29(1):1266--1276, 2023. doi: {{%
10\hspace{.1pt}\discretionary{.}{%
}{.}\hspace{.4pt}1109\discretionary{/}{%
}{/}TVCG\hspace{.1pt}\discretionary{.}{%
}{.}\hspace{.4pt}2022\hspace{.1pt}\discretionary{.}{%
}{.}\hspace{.4pt}3209435}}


\bibitem{wang2019atmseer}
Q.~Wang, Y.~Ming, Z.~Jin, Q.~Shen, D.~Liu, M.~J. Smith, K.~Veeramachaneni, and H.~Qu.
\newblock Atmseer: Increasing transparency and controllability in automated machine learning.
\newblock In {\em Proceedings of the 2019 CHI Conference on Human Factors in Computing Systems}, p. 1–12. Association for Computing Machinery, New York, NY, USA, 2019. doi: {{%
10\hspace{.1pt}\discretionary{.}{%
}{.}\hspace{.4pt}1145\discretionary{/}{%
}{/}3290605\hspace{.1pt}\discretionary{.}{%
}{.}\hspace{.4pt}3300911}}


\bibitem{wang2019genealogy}
Q.~Wang, J.~Yuan, S.~Chen, H.~Su, H.~Qu, and S.~Liu.
\newblock Visual genealogy of deep neural networks.
\newblock {\em IEEE Transactions on Visualization and Computer Graphics}, 26(11):3340--3352, 2020. doi: {{%
10\hspace{.1pt}\discretionary{.}{%
}{.}\hspace{.4pt}1109\discretionary{/}{%
}{/}TVCG\hspace{.1pt}\discretionary{.}{%
}{.}\hspace{.4pt}2019\hspace{.1pt}\discretionary{.}{%
}{.}\hspace{.4pt}2921323}}


\bibitem{yang2020visual}
F.~Yang, Z.~Huang, J.~Scholtz, and D.~L. Arendt.
\newblock How do visual explanations foster end users' appropriate trust in machine learning?
\newblock In {\em Proceedings of the 25th International Conference on Intelligent User Interfaces}, p. 189–201. Association for Computing Machinery, New York, NY, USA, 2020. doi: {{%
10\hspace{.1pt}\discretionary{.}{%
}{.}\hspace{.4pt}1145\discretionary{/}{%
}{/}3377325\hspace{.1pt}\discretionary{.}{%
}{.}\hspace{.4pt}3377480}}


\bibitem{yang2020drift}
W.~Yang, Z.~Li, M.~Liu, Y.~Lu, K.~Cao, R.~Maciejewski, and S.~Liu.
\newblock Diagnosing concept drift with visual analytics.
\newblock In {\em 2020 IEEE Conference on Visual Analytics Science and Technology (VAST)}, pp. 12--23, 2020. doi: {{%
10\hspace{.1pt}\discretionary{.}{%
}{.}\hspace{.4pt}1109\discretionary{/}{%
}{/}VAST50239\hspace{.1pt}\discretionary{.}{%
}{.}\hspace{.4pt}2020\hspace{.1pt}\discretionary{.}{%
}{.}\hspace{.4pt}00007}}


\bibitem{you2024gamifying}
Y.~You, H.~W. Chen, and J.~Zhao.
\newblock Enhancing ai explainability for non-technical users with llm-driven narrative gamification.
\newblock In {\em Proceedings of the Extended Abstracts of the CHI Conference on Human Factors in Computing Systems}. Association for Computing Machinery, New York, NY, USA, 2025. doi: {{%
10\hspace{.1pt}\discretionary{.}{%
}{.}\hspace{.4pt}1145\discretionary{/}{%
}{/}3706599\hspace{.1pt}\discretionary{.}{%
}{.}\hspace{.4pt}3719795}}


\bibitem{yuan2020survey}
J.~Yuan, C.~Chen, W.~Yang, M.~Liu, J.~Xia, and S.~Liu.
\newblock A survey of visual analytics techniques for machine learning.
\newblock {\em Computational Visual Media}, 7(1):3--36, 2021. doi: {{%
10\hspace{.1pt}\discretionary{.}{%
}{.}\hspace{.4pt}1007\discretionary{/}{%
}{/}s41095\discretionary{%
}{-}{-}020\discretionary{%
}{-}{-}0191\discretionary{%
}{-}{-}7}}


\end{thebibliography}
